\documentclass[%
 aip,
 jmp,%
 amsmath,amssymb,
 reprint,%
]{revtex4-1}
\usepackage{color}
\usepackage{graphicx}
\usepackage{dcolumn}
\usepackage{bm}

\begin{document}

\newcommand{\SiNx}{$\mathrm{SiN}_x$}
\newcommand{\Ohm}{$\mathrm{\Omega}$}
\newcommand{\uOhmcm}{$\mathrm{\mu\Omega\ cm}$}
\newcommand{\Ql}{$Q_{\mathrm{l}}$}
\newcommand{\Qc}{$Q_{\mathrm{c}}$}
\newcommand{\Qcone}{$Q_{\mathrm{c1}}$}
\newcommand{\Qctwo}{$Q_{\mathrm{c2}}$}
\newcommand{\Qi}{$Q_{\mathrm{i}}$}
\newcommand{\um}{$\mathrm{\mu m}$}
\newcommand{\Tc}{$T_\mathrm{c}$}
\newcommand{\celcius}{$\,^{\circ}\mathrm{C}$}

\title[Endo {\it et al.}]{On-chip filter bank spectroscopy at 600-700 GHz using NbTiN superconducting resonators}

\author{A. Endo}%
\affiliation{ 
Kavli Institute of NanoScience, Faculty of Applied Sciences, Delft University of Technology, Lorentzweg 1, 2628 CJ Delft, The Netherlands
}%
 \email{A.Endo@tudelft.nl.}
\author{C. Sfiligoj}
\affiliation{ 
Kavli Institute of NanoScience, Faculty of Applied Sciences, Delft University of Technology, Lorentzweg 1, 2628 CJ Delft, The Netherlands
}%
\author{S.J.C. Yates}
\affiliation{ 
SRON, Landleven 12, 9747 AD Groningen, The Netherlands
}%
\author{J.J.A. Baselmans}
\affiliation{ 
SRON, Sorbonnelaan 2, 3584 CA Utrecht, The Netherlands
}%
\author{D.J. Thoen}
\affiliation{ 
Kavli Institute of NanoScience, Faculty of Applied Sciences, Delft University of Technology, Lorentzweg 1, 2628 CJ Delft, The Netherlands
}%
\author{S.M.H. Javadzadeh}
\affiliation{ 
Kavli Institute of NanoScience, Faculty of Applied Sciences, Delft University of Technology, Lorentzweg 1, 2628 CJ Delft, The Netherlands
}%
\author{P.P. van der Werf}
\affiliation{ 
Leiden Observatory, Leiden University, PO Box 9513, NL-2300 RA Leiden, The Netherlands
}%
\author{A.M. Baryshev}
\affiliation{ 
SRON, Landleven 12, 9747 AD Groningen, The Netherlands
}%
\affiliation{ 
Kapteyn Astronomical Institute, University of Groningen, P.O. Box 800, 9700 AV Groningen, The Netherlands
}%
\author{T.M. Klapwijk}
\affiliation{ 
Kavli Institute of NanoScience, Faculty of Applied Sciences, Delft University of Technology, Lorentzweg 1, 2628 CJ Delft, The Netherlands
}%

\date{\today}

\begin{abstract}
We experimentally demonstrate the principle of an on-chip submillimeter wave filter bank spectrometer, using superconducting microresonators as narrow band-separation filters. The filters are made of NbTiN/\SiNx/NbTiN microstrip line resonators, which have a resonance frequency in the range of 614-685 GHz---two orders of magnitude higher in frequency than what is currently studied for use in circuit quantum electrodynamics and photodetectors. The frequency resolution of the filters decreases from 350 to 140 with increasing frequency, most likely limited by dissipation of the resonators.
\end{abstract}
\maketitle


On-chip filter bank spectrometers using superconducting resonators as narrow band-separation filters have been theoretically studied\cite{Endo:2012in, Endo:2012ed, Kovacs:fn} as one of the promising technologies for enabling multi-object broadband spectrometers on next-generation millimeter/submillimeter wave telescopes\cite{Woody:2012fh} for astronomy. The advantage of an integrated filter bank spectrometer, such as the proposed instrument DESHIMA (Delft-SRON High z Mapper)\cite{Endo:2012in, Endo:2012ed}, over conventional optical spectrometers\cite{Stacey:uh} is the combination of; 1) photon-noise limited point-source sensitivity, ultimately equal to a grating spectrometer, 2) compact size of each filter, which is of the order of the wavelength on chip and independent of frequency resolution, and 3) flexibility for increasing the sampling in either frequency space, or in real space, by arranging multiple filter bank units on a focal plane.
One of the challenges towards realization is to make resonant filters with $Q \ge 300$, to match the typical line width of distant submillimeter galaxies\cite{Carilli:2006em}. Unloaded
$Q$'s of up to 2000 have been reported\cite{2009AIPC.1185..164G} at 100 GHz ($\lambda = 3$ mm), but so far there has been no experimental study on resonant filters with $Q > 100$ in the submillimeter band (300-1000 GHz), in which the luminous C$^+$ line at 1.9 THz can be detected from dusty star-forming galaxies in the redshift range of 1$<z<$5, including the peak of cosmic star formation history.\cite{Endo:2012in, Endo:2012ed}

\begin{figure*}
\includegraphics[width=\textwidth]{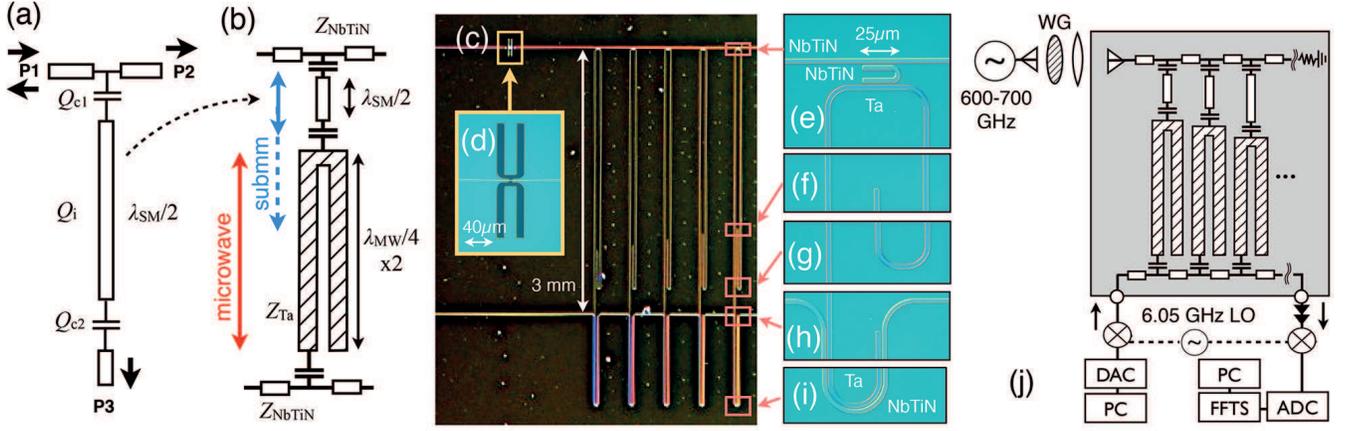}
\caption{\label{fig:chip}
(a) 3-port network model of a single submillimeter wave filter. $\lambda_{\mathrm{SM}}$ and $\lambda_{\mathrm{MW}}$ indicates the submillimeter wave and microwave wavelengths, respectively, which are reduced by the dielectric medium and also by the kinetic inductance of the superconductor. (b) Network model of a single channel, which is a combination of a submillimeter wave filter and a microwave resonator (MKID). The channel is capacitively coupled to the signal line on the filter end, and to the readout line on the MKID end. (c) Micrograph of a small area on the filter bank chip. The image captures the antenna and 5 channels of the filter bank nearest to the antenna. (d) Double-slot antenna etched into the NbTiN ground plane. The horizontal line is the signal MSL which is connected to the filter bank. (e) Submillimeter wave MSL filter, shaped like the character U. (f, g) sections of the MSL-MKID. (h, i) Microwave coupler of the MKID for reading out the response. (j) Block diagram of the experimental setup.
}
\end{figure*}

Here we prove the principle of performing spectroscopy in the 600-700 GHz submillimeter band, using a superconducting on-chip filter bank. 
We have developed a chip, onto which 30 spectroscopic channels are integrated.
A schematic of the distributed circuit is presented in Fig. \ref{fig:chip} (j).
Each channel is a combination of a submillimeter-wave superconducting resonator which functions as a band-separation filter, and a microwave resonator coupled to it which functions as a microwave kinetic inductance detector (MKID), 
and is indicated with a shaded line in Fig. \ref{fig:chip}).
There are two sets of 15 channels on one chip, which are each designed to have a resolution of $f/df = 1000$, and spaced in frequency with an interval of 20.8 GHz so that they sparsely cover the band of 530-830 GHz. 
In the case of a real spectrometer, the filters should be packed densely to cover the band with no gaps.
In the following, we will show that this chip is indeed capable of detecting submillimeter waves at different channels depending on the frequency, and discuss the properties of the resonant filters.

We will begin by describing the working principle of a single band-separation filter, using  transmission line theory.\cite{Pozar:2003we}
The configuration of a single filter is represented by a 3-port transmission line model in Fig. \ref{fig:chip} (a). The submillimeter wave signal enters the circuit from port 1 (we denote port 1 as P1 hereafter, and accordingly for P2 and P3) and flows towards P2. We will call the line between P1 and P2 the `signal line'. Along the signal line is a half wavelength resonator which is capacitively coupled to the line. On the other side, it is capacitively coupled to P3, which we will later connect to a matched MKID detector. The loaded quality factor of the resonator, \Ql, consists of 3 components;
\begin{equation}
1/Q_\mathrm{l} = 1/Q_\mathrm{i} + 1/Q_\mathrm{c1}+1/Q_\mathrm{c2}.
\end{equation}
Here, \Qi\ is the unloaded (internal) quality factor that reflects dissipation loss in the resonator, 
\Qcone\ is the quality factor set by the power that leaks from the resonator to P1 and P2, and
\Qctwo\ is the quality factor set by the power leaking to P3. Hereafter we will assume \Qcone = \Qctwo $\equiv$ \Qc, because we have designed the two capacitors symmetrically to achieve maximum coupling. Using these definitions, the normalised magnitude of power transmitted from P1 to P3 for signal frequencies $f$ in the vicinity of the resonance frequency $f_0$ can be written with a Lorentzian function as 
\begin{equation}\label{eq:S31complex} 
|S_{31}|^2 = \Biggl|\frac{Q_\mathrm{l}}{\sqrt{\frac{2Q_\mathrm{i}^2Q_\mathrm{l}^2}{(Q_\mathrm{i}-Q_\mathrm{l})^2}}(1+2i Q_\mathrm{l}\frac{f-f_0}{f_0})}\Biggr|^2.
\end{equation}
At $f=f_0$, the transmission reaches its maximum of
\begin{equation}
|S_{31}|^2 = \frac{2Q_\mathrm{i}^2}{(Q_{\mathrm{c}}+2Q_{\mathrm{i}})^2}.
\end{equation}


The configuration of a single spectroscopic channel, which is a combination of a filter and an MKID detector, is presented in Fig. \ref{fig:chip} (b).
At the top of the diagram is the submillimeter wave filter that we discussed in the previous paragraph. The signal line and the filters are made of NbTiN, which typically has a gap frequency of $\sim$1.1 THz\cite{Jackson:vx} and should therefore behave as a superconductor for a signal of 600-700 GHz.
Now P3 is connected to the center of an MKID which is open-ended on both sides. 
The MKID is itself a half wave resonator for a microwave readout tone in the range of 6-7 GHz, and is capacitively coupled on one of its open ends to a through-line made of a NbTiN microstrip line for probing the microwave response. 
For the prototype device we have chosen Ta as the material for the top wire, while the ground plane is the same NbTiN film continuing from the filter. 
Inferring from the measured $T_\mathrm{c}$ of the Ta film of 4.4 K and assuming the BCS relation of $2\Delta =3.5k_\mathrm{B}T_\mathrm{c}$, the gap frequency of the Ta film is $\sim$320 GHz, and therefore forms a lossy stripline for the signal at 600-700 GHz.


Micrographs of the circuit are presented in Fig. \ref{fig:chip} (c-i). Fig. \ref{fig:chip} (c) shows the double slot antenna (subfigure d) coupled to the signal line, and part of the filter bank. 
Each channel has a different resonance frequency for the filter and the MKID.
Micrographs with higher magnification of a single channel are presented in subfigures (e-i). There is a NbTiN ground plane under the entire circuit, and there is a l \um-thick layer of \SiNx\ in between the ground plane and the wires. The filters have a U shape, and are designed using simulations with SONNET EM\cite{Inc:tm} so that the single-side \Qc\ is equal to 2000.


The fabrication of the device begins with taking a c-plane sapphire wafer with a thickness of 350 \um\ and a diameter of 100 mm. 
A 300 nm-thick layer of NbTiN is dc-sputter deposited to become the ground plane, and the antenna slots were etched using electron beam lithography and dry etching. Then a 1\um-thick layer of amorphous \SiNx\ is deposited using plasma-enhanced chemical vapour deposition (PECVD) at 300\celcius. We subsequently deposit a 200 nm-thick layer of Ta on top of a 7.5 nm-thick seed layer of Nb, followed by electron-beam lithography and dry etching to define the MKIDs. Finally, we deposit 85 nm of NbTiN and define the signal line, the filters, the readout through line, and via-less bonding pads to connect to the sample box. All wires have a width of 3 \um, except at the feed point of the antenna, where it decreases to 1 \um. The top NbTiN layer has a resistivity of 160 \uOhmcm \ and a superconducting transition temperature of \Tc\ = 14.2 K. 


The microwave losses of the transmission lines have been measured using a vector network analyzer while the chip was cooled to 350 mK using a $^3$He sorption cooler. All 30 Ta/\SiNx/NbTiN microstrip MKIDs have a resonance frequency in the range of 6-7 GHz as designed, and have an unloaded quality factor of \Qi\ = 1-5$\times 10^5$. We have also fabricated a single microwave resonator with a NbTiN wire to measure the microwave losses of the NbTiN/\SiNx/NbTiN microstrip line. This resonator showed an unloaded quality factor of 2.5$\times 10^5$. These values show that the \SiNx\ dielectric has a microwave loss in the range of $\tan \delta = 2$-$10 \times 10^{-6}$, which is comparable to or even lower than the best values found in literature\cite{Mazin:2010ga}. 


We have measured the submillimeter wave frequency-dependent response of the channels using a measurement setup as shown in Fig. \ref{fig:chip} (j).
The filter bank chip is shown in the grey rectangle.
On the back side of the chip, we have glued an elliptical lens made of Si, which has a diameter of 8 mm.
The chip and the lens are cooled down to 250 mK using a $^3$He sorption cooler. The lens looks straight out of the cryostat through a 1.1 THz micro mesh low pass filter\cite{2006SPIE.6275E..26A} and a GORE-TEX infrared blocker. Outside the window is a multiplier-based narrow-band submillimeter source with a tuneable frequency in the range of 600-700 GHz, which shines radiation from a feed horn into the cryostat window. The polarisation of the feed horn is tilted by 45$^\circ$ with respect to the polarisation of the antenna on the chip. By turning the orientation of a linear polarising grid in the optical path, we can make the polarisation of the signal entering the cryostat either parallel or perpendicular to the designed polarisation of the antenna on the chip. 

\begin{figure}[t]
\includegraphics{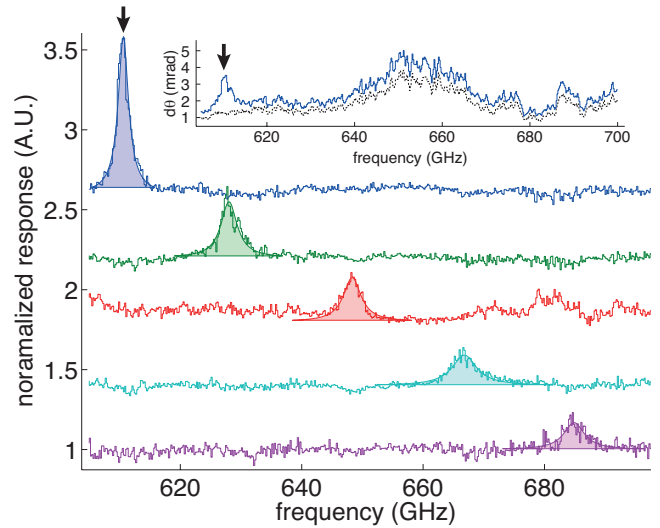}
\caption{\label{fig:spectrum} Normalized phase response of the MKIDs behind concecutive filters in the range of 600-700 GHz. All 5 curves have been normalized so that the average background response outside the filter band is equal to unity, and each curve has been given an offset for clarity. The stepped plots indicate the measured data, where the solid curves are Lorentzian fits to each set of data. 
(Inset) Example of spectra for two channels before dividing out the common stray light component. The vertical axis shows the phase response of the MKIDs in units of milliradians. 
One of the spectra (solid blue curve) has a filter with a frequency within the measured band at 614 GHz, where the other one (dashed black curve) has by design a filter centered at 759 GHz, which is outside the measured frequency band. The spectrum of the 614 GHz channel after dividing out the stray light component is included also in the main figure, as indicated by the arrows.}
\end{figure}

The bandpass characteristics of the filters are measured by observing the response of each MKID while sweeping the submillimeter source frequency.
We use an FFTS-based multi-tone readout electronics\cite{Klein:2012fl, Yates:2009hd} with 400 MHz bandwidth, which allows us to simultaneously measure the response of 16 MKIDs, of which 5 have 1) MKID resonance frequencies within a single readout bandwidth, 2) designed filter resonance frequencies in the 600-700 GHz band, and 3) situated consecutively on the chip.
A common background signal is seen for all 16 MKIDs, 
as shown in the inset of Fig. \ref{fig:spectrum}, where the response of one channel is compared with a trace of a channel with a filter tuned to outside of the 600-700 GHz band.
Because there was no systematic dependence of the common signal on the positions of the resonators on the chip, we conclude that it is due to stray light leaking from around the lens into the sample holder, which forms an integration cavity to couple the stray light directly to the MKIDs.
In a future experiment the amount of stray light can be reduced by a better design of the sample box,\cite{Baselmans:2012gt} and using intermediate optics to improve the coupling efficiency from the source to the lens-antenna.
Here we take advantage of the fact that the background signal is common for all channels, and use it first to calibrate the responsivity variation between the MKIDs.
Then we flatten the spectra by normalizing each spectrum on the background signal. 
As a result, we obtain a spectrum for each channel which has a normalized background level at unity, onto which any response unique to that channel is superimposed.
Fig. \ref{fig:spectrum} shows the obtained spectra for the 5 in-band channels. 
Each spectrum has a single peak, which is fitted well with the theoretically expected Lorentzian curve given in Eq. \ref{eq:S31complex}.
The center frequencies of each filter agree well with the design, with an offset of 4\%.
The center frequencies of the channels increase linearly, with a spacing of $\sim$17 GHz, which is slightly smaller than the designed value of 20.8 GHz. As we rotate the polarizer so that the polarization of the signal from the source is perpendicular to the designed polarization of the antenna, the peak vanishes while the broad band response to stray light remains similar.

\begin{figure}[t]
\includegraphics{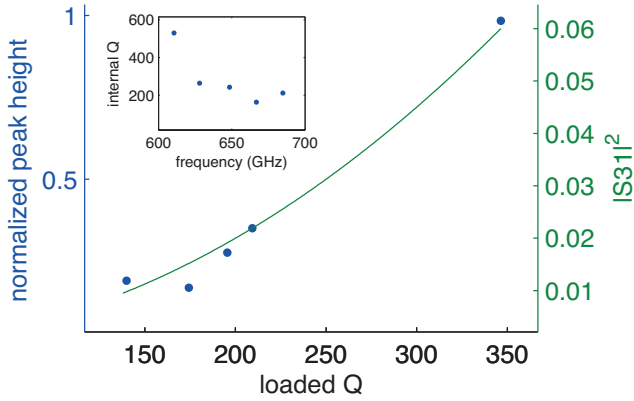}
\caption{\label{fig:fit} 
(Left axis, points) Normalized peak height plotted agianst the loaded quality factor \Ql, where both values are taken from the fitted curves presened in Fig. \ref{fig:spectrum}. (Right axis, curve) Normalized filter transmission $|S_{31}(Q_{\mathrm{l}})|^2$ as a function of \Ql, where both $|S_{31}(Q_{\mathrm{l}})|^2$ and \Ql\ are calculated by varying \Qi\  as a parameter in Eq. \ref{eq:S31complex}. (Inset) Inferred \Qi\  for each filter, plotted against the resonance frequency.
} 
\end{figure}

By fitting each curve in Fig. \ref{fig:spectrum} with Eq. \ref{eq:S31complex}, we extract \Ql\ and the height of the peak above the background level, and plot them against each other in Fig. \ref{fig:fit}.
\Ql\ varies from 140 to 350, while the peak height spans over a factor of 6.
It is interesting to consider whether there is a systematic relation between the two.
In  Fig. \ref{fig:fit}, we have overlaid a parametric plot of $|S_{31}|^2$ as a function of \Ql\  calculated using Eq. \ref{eq:S31complex}, by adopting the designed value of $Q_{\mathrm{c}}=2000$ and sweeping \Qi\ in the range of $160<Q_{\mathrm{i}}<530$.
The reasonably good agreement between the model and the observation indicates that the \Ql\ of the resonators---which sets the resolution of the filter bank as a spectrometer---is limited by dissipation in the resonators, rather than too strong a coupling to the input and output ports.
Note that for the opposite case in which $Q_{\mathrm{c}} < Q_{\mathrm{i}}=\mathrm{const}$ and \Qc\ is swept as a parameter, $|S_{31}(Q_{\mathrm{l}})|^2$ becomes a monotonically  {\it decreasing} curve. From the curve we can tell that the filter transmission $|S_{31}|^2$ is in the range of 1-6\%.
The \Qi\ calculated in this way for each filter is plotted as a function of resonance frequency in the inset of Fig. \ref{fig:fit}. \Qi\ begins at 530 for $f_0=$ 614 GHz, and decreases by a factor of $\sim$3 as the frequency is increased.

If we would now assume that this decrease in \Qi\ with frequency is due to the increase in dielectric loss, it would imply that the loss tangent ($\tan \delta = Q_{\mathrm{i}}^{-1}$) of the \SiNx\ film increases from  $2 \times 10^{-3}$ at 614 GHz to  $6 \times 10^{-3}$ at 685 GHz. Though the absolute magnitude of these values are comparable to the room-temperature losses at the same frequencies reported for CVD-deposited \SiNx\ films\cite{2012OptL...37.4200C}, the losses reported therein increase by only $\sim$15\% over the range of 600-700 GHz. This is strikingly slow compared to the rapid increase by a factor of 3 in our experiment.

If we next assume that the superconducting NbTiN is responsible for the loss, the sheet resistance inferred from the \Qi\ values are in the range of 0.7-2 m\Ohm, 
which is lower than epitaxial NbN and NbCN films measured at 4.2 K\cite{2002PhyC..378.1295K,Kohjiro:tr}.  
If we would treat NbTiN in the framework of the BCS and Mattis-Bardeen theory\cite{Mattis:1958vo} as a superconductor with \Tc\ = 14.2 K and hence a gap energy of 2.2 meV, the \Qi\ of a resonator for 600-700 GHz at a temperature of 250 mK should be many orders of magnitude higher than what can be probed in this experiment. One possibility is that there is a layer of reduced \Tc\ at the surface of the NbTiN, produced either during the 300\celcius-deposition of \SiNx, or the initial stages of the deposition of the wire, as has been argued for Nb strip lines.\cite{2009ApPhL..95y3502Z} 
In order to explain the \Qi\ which we measure, one would have to assume such an interface layer with a \Tc\ of 8K or lower. 

In conclusion, we have experimentally demonstrated submillimeter-wave on-chip spectroscopy using a superconducting filter bank, in the submillimeter wave band of 600-700 GHz. The achieved resolution is in the range of \Ql\ = 140-350, which is found to be limited by losses in the transmission line. The estimated transmission across the filters is 1-6\%, which can be increased by bringing \Qc \ closer to the \Qi\, at the cost of a lower resolution. Development of transmission lines with losses lower by a factor of 4-10 (\Qi $\ge 2000$), sample packaging with better stray light control, and antennae with a high efficiency over a broad bandwidth\cite{Neto:2010vj}, are required for the realization of on-chip filter bank spectrometers with sufficient resolution and photo-efficiency to be useful for extragalactic astronomical science at these high frequencies.

We would like to thank A. Bruno and M. Bruijn for film deposition, and D. Cavallo and L. Ferrari for  advice on the antenna design. AE is financially supported by NWO (Veni grant 639.041.023) and JSPS Fellowship for Research Abroad. This research was partially supported by the NWO Medium Investment grant (614.061.611).


%

\end{document}